\documentstyle[twoside,fleqn,npb,epsfig]{article}
\newcommand{\ep}{\varepsilon}
\newcommand{\Li}[2]{{\mbox{Li}}_{#1}\left(#2\right)}

\newcommand{\Ls}[2]{{\mbox{Ls}}_{#1}\!\left(#2\right)}
\newcommand{\LS}[3]{{\mbox{Ls}}_{#1}^{(#2)}\!\left(#3\right)}

\newcommand{\Ti}[2]{{\mbox{Ti}}_{#1}\left(#2\right)}
\newcommand{\tfrac}[2]{{\textstyle{\frac{#1}{#2}}}}

\newcommand{\Snp}[2]{{\mbox{S}}_{#1\!}\left(#2\right)}

\title{Series and  $\ep$-expansion of the hypergeometric functions}

\author{
M.~Yu.~Kalmykov \address{
JINR, 141980, Dubna (Moscow Region),Russia \\
DESY, Platanenallee 6,D-15738, Zeuthen, Germany}
\thanks{Work was supported by DFG under Contract SFB/TR 9-03 
    and in part by RFBR grant \# 04-02-17192.}
}

\begin{document}

\begin{abstract}
Recent progress in analytical calculation of the 
{\it multiple \{inverse, binomial, harmonic\} sums }, related with 
$\ep$-expansion of the hypergeometric function of one variable are discussed. 
\end{abstract}

\maketitle
{\bf 1.}
In the framework of the dimensional regularization \cite{dimreg} many Feynman
diagrams can be written as hypergeometric series of several variables \cite{feynman-hyper}
(some of them can be equal to the rational numbers).
This result can be deduced via Mellin-Barnes technique \cite{mellin}
or as solution of the differential equation for Feynman amplitude \cite{diff}. 
However, for application to the calculation of radiative corrections
mainly the construction of the $\ep$-expansion is interesting. 
At the present moment, several algorithms for the $\ep$-expansion of 
different hypergeometric functions have been elaborated. They were mainly 
related to calculations of concrete Feynman diagrams \cite{qed}. 
Only recently, the general algorithm for integer values of
parameters has been described in \cite{nested} and its generalization has
been done in \cite{rational}. The results of expansion are expressible in
terms of the new functions, like harmonic polylogarithms \cite{RV00} or their 
recent generalization \cite{nested,generalization}.
Let us shortly describe, how this algorithm woks on the example 
of the generalized hypergeometric function of one variable. The starting point 
is the series representation: 
\vspace*{-1.5mm}
\begin{eqnarray}
&& \hspace*{-7mm}
_{P}F_Q\left(\begin{array}{c|}
\!\!\! \{A_1\!+\!a_1\ep\}, \!\! \{A_2\!+\!a_2\ep\},  \cdots \{A_P\!+\!a_P\ep\}  \\
\!\!\! \{B_1\!+\!b_1\ep\}, \!\! \{B_2\!+\!b_2\ep\},  \cdots \{B_Q\!+\!b_q\ep\}
\end{array} ~z \right)
\nonumber \\ && \hspace*{-3mm}
= 
\sum_{j=0}^{\infty} \frac{z^j}{j!}\;
\frac{\Pi_{s=1}^P(A_s+a_s\ep)_j}{\Pi_{r=1}^Q(B_r+b_r\ep)_j} \; ,
\nonumber 
\end{eqnarray}
where 
$(\alpha)_j\equiv\Gamma(\alpha+j)/\Gamma(\alpha)$ 
is the Pochhammer symbol.
We concentrate on the case $_{Q+1}F_Q$, when 
series converges for all $|z|<1$,
and on the integer or  half-integer values of the parameters
$\{ A_i, \;  B_j \} \in \{m_i, m_j+\frac{1}{2} \}$.
To perform the $\ep$-expansion we use the well-known representation
\vspace*{-3mm}
\begin{eqnarray}
&&  \hspace{-7mm}
\frac{(m+a\ep)_j}{(m)_j} =
\nonumber \\ &&  \hspace{-7mm}
\exp\left\{ -\sum_{k=1}^{\infty} \frac{(-a\ep)^k}{k}
\left[ S_k(m \!+\! j \!-\! 1) \!-\! S_k(m \!-\! 1) \right] \right\} \; ,
\nonumber 
\end{eqnarray}
where $m$ is an integer positive number, $m > 1$
and 
$S_k(j) = \sum_{l=1}^j l^{-k}$ 
is the harmonic sum satisfying the relation
$
S_k(j) = S_k(j-1) + 1/j^k \;. 
$
For half-integer positive values ${\cal A}_i \equiv m_i + 1/2 >0$,
we use the duplication formula
$$
\left( m + \frac{1}{2} + a \ep \right)_{j}=
\frac{\left( 2 m + 1 + 2 a \ep \right)_{2 j} }{4^j \left( m + 1 + a \ep \right)_{j}} \; . 
$$
To work only with positive values for parameters of hypergeometric function
we can apply several times the Kummer relation:
\begin{eqnarray}
&& \hspace{-8mm}
{}_{P}F_Q \left(\begin{array}{c|} a_1, \cdots, a_P \\ 
b_1, \cdots, b_Q \end{array} ~z \right) = 1 
\nonumber \\ && \hspace{-8mm}
\!+\! z \frac{a_1 \ldots a_P}{b_1 \ldots b_Q}  
~{}_{P\!+\!1}F_{Q\!+\!1} \left(\begin{array}{c|} 1, 1 \!+\!a_1,\cdots, 1 \!+\!a_P \\ 
2, 1 \!+\! b_1,\cdots, 1 \!+\! b_Q \end{array} ~z \right) \; . 
\nonumber 
\end{eqnarray}
After applying this procedure the original hypergeometric function can be written as 
\begin{eqnarray}
&& \hspace*{-7mm}
{}_{P\!+\!1}F_P \left(\begin{array}{c|} 
\{ m_i\!+\!a_i\ep \}^{J}, \{ p_j\!+\! \tfrac{1}{2} \!+\! d_j \ep \}^{P\!+\!1\!-\!J} \\ 
\{ n_i\!+\!b_i\ep \}^K,  \{ l_j\!+\! \tfrac{1}{2} \!+\! c_j \ep \}^{P\!-\!K}  
\end{array} ~z \right) = 
\nonumber \\ && \hspace*{-3mm}
\sum_{j=1}^\infty \frac{z^j}{j!} \frac{1}{4^{j(K\!-\!J\!+\!1)}} 
\frac{\Pi_{i=1}^J (m_i)_j}{\Pi_{l=1}^K (n_l)_j}
\nonumber \\ && 
\times
\Pi_{r=1}^{P\!+\!1\!-\!J} \frac{(2 p_r\!+\!1)_{2j}}{(p_r\!+\!1)_{j}}
\Pi_{s=1}^{P\!-\!K} \frac{(l_s\!+\!1)_j}{(2 l_s\!+\!1)_{2j}}
\Delta \; , 
\label{hyper}
\end{eqnarray}
\vspace{-5mm}
with
\begin{eqnarray}
&& 
\Delta  =  
\exp \Biggl[ \sum_{k=1}^{\infty} \frac{(-\ep)^k}{k}  
\Biggl( 
\sum_{\omega=1}^K b_\omega^k S_k(n_\omega+j-1)
\nonumber \\[-2mm] && 
- \sum_{i=1}^J a_i^k S_k(m_i+j-1)
\nonumber \\[-1mm] && 
+ \sum_{s=1}^{P-K} c_s^k \left[ S_k(2 l_s+2j) - S_k(l_s+j) \right]
\nonumber \\[-1mm] &&
- \sum_{r=1}^{P+1-J} d_r^k \left[ S_k(2 p_r+2j) - S_k(p_r+j) \right]
\Biggr) 
\Biggr] \; . 
\nonumber 
\end{eqnarray}
In this way, the $\ep$-expansion of the hypergeometric function 
(\ref{hyper}) is reduced to the calculation of the {\it multiple sums}
\vspace*{-1.5mm}
\begin{eqnarray}
&& \hspace{-7mm}
\sum_{j=1}^\infty \frac{z^j}{j!} \frac{1}{4^{j(K\!-\!J\!+\!1)}} 
\Pi_{i,r,\omega,s}
\frac{ (m_i-1+j)! (2 p_r+2j)!}{(n_\omega-1+j)! (2 l_s+2j)!} 
\nonumber \\[-2mm] && \hspace{-5mm}
\times [S_{a_1}(m_1\!+\!j\!-\!1)]^{i_1}\ldots [S_{a_\mu}(m_\mu\!+\!j\!-\!1)]^{i_p}\; 
\nonumber \\ && \hspace{-5mm}
\times [S_{b_1}(2p_r\!+\!2j)]^{j_1}\ldots [S_{b_\nu}(2p_\nu\!+\!2j)]^{j_q},
\nonumber 
\end{eqnarray}
where $\{m_j,n_k,l_\omega,p_r\}$ - positive integer numbers.
In the calculation of massive Feynman diagrams \cite{diagrams,DK01,DK04} we get  
{\it multiple sums} of the following form, 
\vspace*{-3mm}
\begin{eqnarray}
&& 
\Sigma_{a_1,\ldots,a_p; \; b_1,\ldots,b_q;c}(u)
\equiv
\sum_{j=1}^\infty \tfrac{1}{\left( 2j \atop j\right)^k }\frac{u^j}{j^c}
\nonumber \\[-2mm] &&
\times
S_{a_1} \ldots S_{a_p} \bar{S}_{b_1} \ldots \bar{S}_{b_q} \; , 
\label{binsum}
\end{eqnarray}
where $u$ is an arbitrary argument
and we accept that the notation $S_a$ and $\bar{S}_b$
will always mean $S_a(j-1)$ and $S_b(2j-1)$, respectively, 
even we do not mention this explicitly.
When there are no sums of the type $S_{a}$ or $\bar{S}_{b}$
in the r.h.s. of Eq.~(\ref{binsum}),
we put a ``$-$'' sign instead of the indices $(a)$ or $(b)$
of $\Sigma$, respectively. Some indices $(a)$ or $(b)$ 
may be equal to each other, which is equivalent to power of a proper harmonic sum.
For particular values of $k$, the sums (\ref{binsum}) are called 
\begin{eqnarray}
k = 
\left\{ 
\begin{array}{cl}
 0  & \mbox{ {\it harmonic} } \\
 1  & \mbox{ {\it inverse binomial} } \\
-1  & \mbox{ {\it binomial} } 
\end{array} \right\} \mbox{ sums }
\nonumber 
\end{eqnarray}
These sums are related to $\ep$-expansion of the hypergeometric 
functions of type (\ref{hyper}) with the following set of parameters: 
\begin{eqnarray}
&&  \hspace*{-7mm}
m_i \in \{ 1 \}^K, \; \{ 2 \}^L  \;,           
n_i \in \{ 1 \}^R, \; \{ 2 \}^{K+L-R-1-k}  \;, 
\nonumber \\ &&  \hspace*{-7mm}
p_j \in \{ 1 \}^{J-k}  \;,  \quad l_j \in \{ 1 \}^{J}, \quad
u = 4^k z \;.  
\end{eqnarray}
In the recent paper \cite{DK04}, the sums of type (\ref{binsum})
up to weight ${\bf 4}$ have been studied in detail.

\noindent
{\bf 2.}
Let us rewrite the {\it multiple sums} (\ref{binsum}) in the following form, 
$
 \Sigma^{(k)}_{A;B;c}(u) = \sum_{j=1}^{\infty} u^j \eta^{(k)}_{A;B;c}(j) \; , 
$
where $A\equiv \left( a_1,\ldots,a_p\right)$
and $B\equiv \left(b_1,\ldots,b_q\right)$
denote the collective sets of indices, whereas $\eta^{(k)}_{A;B;c}(j)$ is
the coefficient of $u^j$
\vspace*{-3mm}
\begin{eqnarray}
&& \hspace{-5mm}
\eta^{(k)}_{A;B;c}(j) = 
\tfrac{1}{\left( 2j \atop j\right)^k }\frac{1}{j^c}
S_{a_1} \ldots S_{a_p} \bar{S}_{b_1} \ldots \bar{S}_{b_q} \; . 
\label{coeff}
\end{eqnarray}
The idea is to find a recurrence relation
with respect to~$j$, for the coefficients $\eta^{(k)}_{A;B;c}(j)$ and 
then transform it into a differential equation
for the {\em generating} function $\Sigma^{(k)}_{A;B;c}(u)$.
In this way, the problem of summing the series would be reduced to 
solving a differential equation \cite{wilf}.
Using the explicit form of $\eta^{(k)}_{A;B;c}(j)$ 
given in Eq.~(\ref{coeff}),
the recurrence relation can be written in the following form:
\vspace*{-2mm}
\begin{eqnarray}
&& 
\bigl[ 2 (2j+1) \bigr]^k (j+1)^{c-k} \eta^{(k)}_{A;B;c}(j+1) 
\nonumber \\ && 
= j^c \eta^{(k)}_{A;B;c}(j) + r^{(k)}_{A;B}(j) \; ,
\label{rec:relation}
\end{eqnarray}
where the explicit form of the ``remainder'' $r^{(k)}_{A;B}(j)$ is given by
\vspace*{-3mm}
\begin{eqnarray}
&& \hspace{-7mm} 
{\textstyle \left( 2j \atop j\right)^k} \;
r^{(k)}_{A;B}(j) =
\prod\limits_{k=1}^p \prod\limits_{l=1}^q 
\Biggl\{ \left[ S_{a_k} \!+\! j^{-a_k}\right]^{i_k} \times
\nonumber \\[-2mm] && \hspace{-7mm}
\left[ \bar{S}_{b_l} \!+\! (2j)^{-b_l} \!+\! (2j\!+\!1)^{-b_l} \right]^{j_l}
\!-\! \left[ S_{a_k} \right]^{i_k} \left[ \bar{S}_{b_l} \right]^{j_l} 
\Biggr\}
\; .
\label{r:AB}
\end{eqnarray}
In other words, it contains all contributions generated by
$j^{-a_k}$, $(2j)^{-b_l}$ and $(2j+1)^{-b_l}$ which appear
because of the shift of the index~$j$. 
Multiplying both sides of Eq.~(\ref{rec:relation}) by $u^j$,
summing from 1 to infinity, and using the fact that any extra power
of~$j$ corresponds to the derivative $u({\rm d}/{\rm d}u)$,
we arrive at the following differential equation for the
generating function $\Sigma^{(k)}_{A;B;c}(u)$: 
\vspace*{-3mm}
\begin{eqnarray}
&& \hspace{-7mm}
\left[ 
\left(\tfrac{4}{u} \!-\! 1 \right) u \tfrac{{\rm d}}{{\rm d} u} 
\!-\! \tfrac{2}{u} 
\right]
\left( u \tfrac{{\rm d}}{{\rm d}u} \right)^{c-1} 
\Sigma^{(1)}_{A;B;c}(u)  
\nonumber \\ && \hspace{-5mm}
= \delta_{p0} + R^{(1)}_{A;B}(u) \; ,
\label{diff:I}
\\ && \hspace{-7mm}
\left(\tfrac{1}{u} \!-\! 1 \right) \left( u \tfrac{{\rm d}}{{\rm d} u} \right)^c  
\Sigma^{(0)}_{A;B;c}(u) =
\delta_{p0} \!+\! R^{(0)}_{A;B}(u) \; ,
\label{diff:II}
\\ && \hspace{-7mm}
\left[ 
\left(\tfrac{1}{u} \!-\! 4 \right) u \tfrac{{\rm d}}{{\rm d} u}
\!-\! 2 
\right]
\left( u \tfrac{{\rm d}}{{\rm d}u} \right)^{c} 
\Sigma^{(-1)}_{A;B;c}(u)  
\nonumber \\ && \hspace{-5mm}
= 2 \delta_{p0} + 2 \left( 2 u \tfrac{{\rm d}}{{\rm d} u} + 1\right) R^{(-1)}_{A;B}(u) \; ,
\label{diff:III}
\end{eqnarray}
where $R^{(k)}_{A;B}(u)\equiv\sum_{j=1}^{\infty} u^j r^{(k)}_{A;B}(j)$.
The r.h.s. of differential equation for sums 
includes the {\it multiple sums} with shifted index
\vspace*{-2mm}
\begin{eqnarray}
&& \hspace{-7mm}
G^{(k)}_{a_1,\ldots,a_p;\; b_1,\ldots,b_q;c}
\equiv
\sum_{j=1}^\infty \tfrac{1}{\left( 2j \atop j\right)^k } \frac{u^j}{(2j+1)^c}
\nonumber \\ && \hspace{-7mm}
\times
S_{a_1} \ldots S_{a_p} \bar{S}_{b_1} \ldots \bar{S}_{b_q} 
\equiv
\sum_{j=1}^{\infty} u^j \nu^{(k)}_{A;B;c}(j) \; , 
\label{binsumII}
\end{eqnarray}
where we accept the same conditions for the indices $\{a_i\}$ and $\{b_j\}$,
as in the previous case. 
For investigation of these sums we again apply the generating function approach. 
In this case, the recurrence relations for the coefficients 
$\nu^{(k)}_{A;B;c}(j)$ are the following:
\vspace*{-2mm}
\begin{eqnarray}
&& \hspace{-5mm}
\bigl[ 2 (2j+1) \bigr]^k (2j+3)^c \nu^{(k)}_{A;B;c}(j+1) 
\nonumber \\ && \hspace{-5mm}
= (j+1)^k \left[ (2j+1)^c \nu^{(k)}_{A;B;c}(j) + r^{(k)}_{A;B}(j) \right] \; ,
\end{eqnarray}
with  $r^{(k)}_{A;B}(j)$ given by Eq.~(\ref{r:AB}).
The proper set of differential equations is
\vspace*{-1.5mm}
\begin{eqnarray}
&& \hspace{-7mm}
\left[
\left(\tfrac{4}{u} \!-\! 1 \right) u \tfrac{{\rm d}}{{\rm d} u} 
\!-\! \tfrac{2}{u} 
\!-\! 1 
\right]
\left( 2 u \tfrac{{\rm d}}{{\rm d} u} + 1\right)^c  
G^{(1)}_{A;B;c}(u)  
\nonumber \\ && \hspace{-5mm}
= \delta_{p0} + \left( u \tfrac{{\rm d}}{{\rm d} u} + 1\right) R^{(1)}_{A;B}(u) \; ,
\label{diff:IB}
\\ && \hspace{-7mm}
\left(\tfrac{1}{u} \!-\! 1 \right) \left( 2 u \tfrac{{\rm d}}{{\rm d} u} + 1\right)^c  
G^{(0)}_{A;B;c}(u) =
\nonumber \\ && \hspace{-5mm}
\delta_{p0} \!+\! \left( u \tfrac{{\rm d}}{{\rm d} u} + 1\right) R^{(0)}_{A;B}(u) \; ,
\label{diff:IIB}
\\ && \hspace{-7mm}
\left[ 
\left(\tfrac{1}{u} \!-\! 4 \right) u \tfrac{{\rm d}}{{\rm d} u} 
\!-\! 2 
\right]
\left( 2 u \tfrac{{\rm d}}{{\rm d}u} + 1\right)^{c} 
G^{(-1)}_{A;B;c}(u)  
\nonumber \\ && \hspace{-5mm}
= 2 \delta_{p0} + 2 \left( 2 u \tfrac{{\rm d}}{{\rm d} u} + 1\right) R^{(-1)}_{A;B}(u) \; .
\label{diff:IIIB}
\end{eqnarray}
Equations, 
$[(\ref{diff:I}),(\ref{diff:IB})]$,
$[(\ref{diff:II}),(\ref{diff:IIB})]$,
$[(\ref{diff:III}),(\ref{diff:IIIB})]$,
form the closed system of differential equations.

\noindent
{\bf 3.}
From the analysis, given in \cite{DK04} 
we have deduced that the set of equations for {\it generating functions}
has a simpler form in terms of a new variable. 
For {\it multiple inverse binomial sums} it is defined as 
\vspace*{-1.5mm}
\begin{equation}
y = \frac{\sqrt{u-4}-\sqrt{u}}{\sqrt{u-4}+\sqrt{u}} \;, \quad
u = - \frac{(1-y)^2}{y} \;, \quad  
\end{equation}
and for {\it multiple binomial sums} it has the following form: 
\vspace*{-1.5mm}
\begin{equation}
\chi = \frac{1-\sqrt{1-4u}}{1+\sqrt{1-4u}}, \quad 
u = \frac{\chi}{(1+\chi)^2}.
\end{equation}
Let us consider the differential equation for {\it multiple inverse binomial sums} 
in terms of new variables. Equation (\ref{diff:I}) takes the form
\vspace*{-2mm}
\begin{eqnarray}
&& \hspace{-5mm}
\left( \!-\! \frac{1\!-\!y}{1\!+\!y} y \frac{d}{d y} \right)^{c-1} 
\Sigma^{(1)}_{A;B;c}(y) \!=\! \frac{1\!-\!y}{1\!+\!y} \sigma^{(1)}_{A;B}(y) \; ,
\label{SIGMA+1}
\end{eqnarray}
where 
\vspace*{-2mm}
\begin{eqnarray}
&& \hspace{-5mm}
y \frac{d}{d y} \sigma^{(1)}_{A;B} \!=\!\delta_{p0} \!+\! R^{(1)}_{A;B}(y) \; .
\end{eqnarray}
Equation (\ref{SIGMA+1}) could be rewritten as 
\vspace*{-2mm}
\begin{equation}
\left( \!-\! \frac{1\!-\!y}{1\!+\!y} y \frac{d}{d y} \right)^{c-j} \Sigma^{(1)}_{A;B;c}(y)  
= \Sigma^{(1)}_{A;B;j}(y) \;,
\end{equation}
or, in equivalent form:
\vspace*{-2mm}
\begin{eqnarray}
&& 
\left( - \frac{1-y}{1+y} y \frac{d}{d y} \right)^{c-j-1} \Sigma^{(1)}_{A;B;c}(y)  
\nonumber \\ && 
= \int_0^y dy \left( \frac{2}{1-y} -  \frac{1}{y} \right) \Sigma^{(1)}_{A;B;j}(y) \;.
\end{eqnarray}
From this representation we immediately get the following {\it statement}: 

\noindent 
If for some $j$ the series $\Sigma^{(1)}_{A;B;j}(y)$
are expressible in terms of {\it harmonic polylogarithms}, the sums 
 $\Sigma^{(1)}_{A;B;j+i}(y)$ can also be presented in terms of 
{\it harmonic polylogarithms}. 
This follows from the definition of the {\it harmonic polylogarithms} (see Ref.~\cite{RV00}).

In a similar manner, let us rewrite the equation for generating function 
of  the {\it multiple binomial sums}: 
\vspace*{-1.5mm}
\begin{eqnarray}
&& \hspace{-6mm}
\left( \frac{1\!+\!\chi}{1\!-\!\chi} \chi \frac{{\rm d}}{{\rm d} \chi}  \right)^c  
\Sigma^{(-1)}_{A;B;c}(\chi)
= \frac{1\!+\! \chi}{1\!-\! \chi} \sigma^{(-1)}_{A;B}(\chi) \; , 
\label{binsumdifI}
\\ && \hspace{-6mm}
\frac{1}{2} (1\!+\!\chi)^2 \frac{{\rm d}}{{\rm d} \chi} \sigma^{(-1)}_{A;B}(\chi) 
\nonumber \\ && \hspace{6mm}
\!=\! 
\delta_{p0} \!+\! 
\left(2 \frac{1\!+\! \chi}{1\!-\! \chi} \chi \frac{{\rm d}}{{\rm d} \chi} \!+\! 1\right) R^{(-1)}_{A;B}(\chi) \; ,
\label{binsumdifII}
\end{eqnarray}
which could also be rewritten as 
\vspace*{-1.5mm}
\begin{equation}
\left( \frac{1\!+\! \chi}{1\!-\! \chi} \chi \frac{{\rm d}}{{\rm d} \chi} \right)^{c-j} \Sigma^{(-1)}_{A;B;c}(\chi)  
= \Sigma^{(-1)}_{A;B;j}(\chi) \;,
\end{equation}
or, in an equivalent form:
\vspace*{-3mm}
\begin{eqnarray}
&& 
\left( \frac{1\!+\! \chi}{1\!-\! \chi} \chi \frac{{\rm d}}{{\rm d} \chi} \right)^{c-j-1} \Sigma^{(-1)}_{A;B;c}(\chi)  
\nonumber \\ && 
= \int_0^\chi d \chi \left( \frac{1}{\chi} - \frac{2}{1+\chi} \right) \Sigma^{(1)}_{A;B;j}(\chi) \;.
\end{eqnarray}
Again we get the previous statement. 

\noindent
{\bf 4.}
The differential equation for {\it multiple inverse binomial sums} with the 
shifted index has a more complicated form. 
For their analysis let us use the geometrical variable \cite{DD}
defined via $u_\theta \equiv 4 \sin^2 \frac{\theta}{2}$ $(0 \leq u_\theta \leq 4)$.
In terms of this variable, Eq.~(\ref{diff:IB}) could be written as 
\vspace*{-1.5mm}
\begin{eqnarray}
&& \hspace{-7mm}
\left[ \cot \tfrac{\theta}{2} \tfrac{d}{d \theta} 
       \!-\! \tfrac{1}{2 \sin^2 \tfrac{\theta}{2}} \!-\! 1 
\right]
\left( 2 \tan \tfrac{\theta}{2}  \tfrac{d}{d \theta} \!+\! 1 \right)^c 
G^{(1)}_{A;B;c}(u_\theta)
\nonumber \\ && \hspace{-7mm}
= 
\delta_{p0} \!+\! 
\left( 1 \!+\! \tan \tfrac{\theta}{2} \tfrac{d}{d \theta} \right) R^{(1)}_{A;B}(u_\theta) \; .
\end{eqnarray}
This equation can be decomposed into the system of differential equations
\vspace*{-1.5mm}
\begin{eqnarray}
&&  \hspace{-7mm}
G^{(1)}_{A;B;c}(u_\theta) = \frac{1}{\sin \frac{\theta}{2}} \rho_{A;B;c} (\theta)\;, 
\label{diff:shifted:1}
\\ &&  \hspace{-7mm}
\left( 2 \tan \tfrac{\theta}{2} \frac{d}{d \theta} \right)^c
\rho_{A;B;c}(\theta) 
=  \frac{ \sin^2 \tfrac{\theta}{2}}{\cos^3 \tfrac{\theta}{2}} g_{A;B}(\theta) \;, 
\label{diff:shifted:2}
\\ &&  \hspace{-7mm}
\tan \tfrac{\theta}{2} \frac{d g_{A;B}(\theta)}{ d \theta} \!=\! 
\nonumber \\ &&  
\frac{d}{d \theta}  
\left( \sin^2 \tfrac{\theta}{2}  \; R^{(1)}_{A;B}(u_\theta)
\!-\! \delta_{p0} \tfrac{1}{2} \cos \theta
\right)\;.
\label{diff:shifted:3}
\end{eqnarray}
The formal solution of Eq.~(\ref{diff:shifted:3}) is
\vspace*{-2mm}
\begin{eqnarray}
&& \hspace{-7mm}
g_{A;B;c}(\theta) =  
\tfrac{1}{2} \sin \theta  R^{(1)}_{A;B} (u_\theta)
\nonumber \\ && \hspace{-7mm}
\!+\! \tfrac{1}{2} \int_0^\theta d \phi R^{(1)}_{A;B} (u_\phi)
\!+\! \tfrac{1}{2} \delta_{p0} \left( \theta + \sin \theta \right) \; , 
\end{eqnarray}
where $u_\phi = 4 \sin^2 \tfrac{\phi}{2}$.
Substituting this result in Eq.~(\ref{diff:shifted:2}) and integrating one time we get 
\vspace*{-2mm}
\begin{eqnarray}
&& \hspace{-7mm}
\left( 2 \tan \tfrac{\theta}{2} \frac{d}{d \theta} \right)^{c-1}
\rho_{A;B;c}(\theta) 
\!=\! 
\!-\! \sin \tfrac{\theta}{2} R^{(1)}_{A;B} (u_\theta)
\nonumber \\ && \hspace{-7mm}
\!+\! \frac{1}{2 \cos \tfrac{\theta}{2}}  \int_0^\theta d \phi R^{(1)}_{A;B}(u_\phi)
\!+\!  \delta_{p0}
\left( \frac{1}{2} \frac{\theta}{\cos \tfrac{\theta}{2}} \!-\! \sin \tfrac{\theta}{2}\right) 
\nonumber \\ && \hspace{-7mm}
\!+\!  \int_0^\theta d \phi \sin \frac{\phi}{2} \frac{d R^{(1)}_{A;B}(u_\phi)}{d \phi} \; .
\label{rho}
\end{eqnarray}
For $c=1$ the r.h.s. of Eq.~(\ref{rho}) divided by $\sin \tfrac{\theta}{2}$ 
is the solution for the sum $G^{(1)}_{A;B;1}(u_\theta)$.
Let us now apply this approach to the sum $G^{(1)}_{-;-;3}(u_\theta)$
which was not solved explicitly in Ref.~\cite{DK04}.
Using the relation (see Eqs.~(\ref{diff:shifted:2}))
\vspace{-2mm}
\begin{equation}
\left( 2 \tan \tfrac{\theta}{2} \frac{d}{d \theta} \right)^{c-k}
\rho_{A;B;c}(\theta) =  
\rho_{A;B;k}(\theta) \;, 
\end{equation}
and expression for $\rho_{-;-;2} (\theta)$ (see Eq.~(2.74) in \cite{DK04}),
$
\rho_{-;-;2} (\theta) = 2 \Ti{2}{\tan \tfrac{\theta}{4}} - \sin \tfrac{\theta}{2} \;, 
$ 
we get after integration by parts 
\vspace{-2mm}
\begin{eqnarray}
&& \hspace{-7mm}
\rho_{-;-;3} (\theta)  = 
\frac{1}{2}
\int_0^\theta d \phi \cot \tfrac{\phi}{2} \rho_{-;-;2} (\phi) 
\nonumber \\ && \hspace{-7mm}
= l_\theta \left[ \rho_{-;-;2} (\theta)  \!+\! \sin \tfrac{\theta}{2} \right]
\!-\! \sin \tfrac{\theta}{2}
\!-\! \tfrac{\theta}{4} \left[ l^2_\theta \!-\! L^2_\theta\right] 
\nonumber \\ && \hspace{-7mm}
\!-\! \frac{1}{2} \left[ 
\Ls{3}{\tfrac{\theta}{2}}
\!+\! \Ls{3}{\pi - \tfrac{\theta}{2}}
\!-\! \Ls{3}{\pi}
\right]
\!-\! \frac{1}{2} I \left( \tfrac{\theta}{2} \right) \; ,
\label{rho3}
\end{eqnarray}
where we have used the short-hand notation
\vspace{-2mm}
$$
l_\theta = \ln \left( 2 \sin \tfrac{\theta}{2}\right) \; , \quad
L_\theta = \ln \left( 2 \cos \tfrac{\theta}{2}\right) \; , 
$$
the integral $I(\theta)$ is defined as
\vspace{-2mm}
\begin{eqnarray}
I(\theta) = \int_0^\theta d \phi \phi \left[ 
   l_\phi \tan \tfrac{\phi}{2} \!+\!  L_\phi \cot \tfrac{\phi}{2}
\right] \;,
\label{NewInt}
\end{eqnarray}
and 
\vspace{-4mm}
\begin{equation}
\label{def_Ls}
\LS{j}{k}{\theta} =   - \int\limits_0^\theta {\rm d}\phi \;
   \phi^k \ln^{j-k-1} \left| 2\sin\frac{\phi}{2}\right| \, ,
\end{equation}
is the generalized log-sine function~\cite{Lewin}. In particular, 
$\LS{j}{0}{\theta} = \Ls{j}{\theta}$.
The integral (\ref{NewInt}) can be evaluated in terms of the polylogarithms of the complex argument \cite{Lewin}
\vspace{-1.5mm}
\begin{eqnarray}
&& \hspace{-7mm}
I(\theta) =  - \tfrac{3}{2} \pi \zeta_2 
\!-\! 3 \Ls{3}{\pi - \theta}
\!-\! \tfrac{1}{2} \Ls{3}{2 \theta}
\!+\! \Ls{3}{\theta}
\nonumber \\ && \hspace{-7mm}
\!+\! 4 \ln 2 \left[\Ls{2}{\pi \!-\! \theta} \!+\! \Ls{2}{\theta} \right]
\!+\! 8 \Im \Snp{1,2}{  i \tan \tfrac{\theta}{2}} 
\nonumber \\ && \hspace{-7mm}
\!+\! 2 \theta \ln \left( \tan \tfrac{\theta}{2} \right)
\left[ \ln 2 \!-\!  \ln \left( \cos \tfrac{\theta}{2} \right) \right] \; , 
\label{I}
\end{eqnarray}
where 
\vspace{-2mm}
\begin{eqnarray}
&& \hspace{-7mm}
\Im \Snp{1,2}{i \tan \tfrac{\theta}{2}}
= 
 \Ti{3}{\tan \tfrac{\theta}{2}, \cot \tfrac{\theta}{2} }
\nonumber \\ && \hspace{-7mm}
\!+\! \tfrac{1}{2} \ln \left( \cos \tfrac{\theta}{2} \right)
\left[ \Ls{2}{\pi \!-\! \theta} 
\!+\! \Ls{2}{\theta} 
\!+\! \theta \ln \left( \tan \tfrac{\theta}{2} \right)
\right] 
\nonumber \\ && \hspace{-7mm}
\!-\! \tfrac{\theta}{8} \Li{2}{\sin^2 \tfrac{\theta}{2}}
\!-\! \tfrac{1}{48} \theta^3 \; ,
\label{ImS12}
\end{eqnarray}
and $\Ti{3}{x,a}$ is the inverse tangent integral of two variables.
Collecting all expressions we get 
\vspace{-1.5mm}
\begin{eqnarray}
&& \hspace{-7mm}
\rho_{-;-;3}(\theta) = 
\tfrac{\theta}{4} \Li{2}{\sin^2 \tfrac{\theta}{4}}
\!-\! 4 \Ti{3}{\tan \tfrac{\theta}{4}, \cot \tfrac{\theta}{4} }
\nonumber \\ && \hspace{-7mm}
\!+\! \Ls{3}{\pi \!-\! \tfrac{\theta}{2}}
\!-\! \Ls{3}{\tfrac{\theta}{2}}
\!+\! \tfrac{1}{4} \Ls{3}{\theta}
\!-\! \Ls{3}{\pi}
\nonumber \\ && \hspace{-7mm}
\!+\! \ln \left( \tan \tfrac{\theta}{4} \right) 
       \left[ \Ls{2}{\tfrac{\theta}{2}} \!+\!  \Ls{2}{\pi- \tfrac{\theta}{2}} \right]
\!+\! \tfrac{\theta^3}{96}
\!-\! \sin \tfrac{\theta}{2}
\nonumber \\ && \hspace{-7mm}
\!+\! \theta \left[ 
  \tfrac{1}{4} \ln^2 \left( \tan \tfrac{\theta}{4} \right)
\!-\! \ln \left( 2 \sin \tfrac{\theta}{4} \right) \ln \left( 2 \cos \tfrac{\theta}{4} \right)
\right]
\nonumber \\ && \hspace{-7mm}
\!+\! \tfrac{\theta}{4} L_\theta \left[ 2 \ln \left( \tan \tfrac{\theta}{4} \right) \!+\! L_\theta  \right] \; . 
\label{rho3:res}
\end{eqnarray}
Combining this result with Eqs.~(2.61), (2.62) and (2.54) from Ref.~\cite{DK04}
we write the one-fold integral  representation for the sum $\Sigma_{;-;3;1}^{(1)}(u_\theta)$
\vspace{-3mm}
\begin{eqnarray}
&& \hspace{-7mm}
\Sigma^{(1)}_{-;3;1}(u_{\theta}) =  \frac{1}{8} \; \tan\tfrac{\theta}{2}\;
\left[2 \theta \LS{3}{1}{\theta} - 3 \LS{4}{2}{\theta} \right]
\nonumber \\ && \hspace{-7mm}
+ \tan\tfrac{\theta}{2}\; \int_0^\theta d \phi \left(1 + \frac{\rho_{-;-;3}(\phi)}{\sin \tfrac{\phi}{2}} \right) 
\;, 
\label{SIGMA3}
\end{eqnarray}
where $\LS{3}{1}{\theta}$ and $\LS{4}{2}{\theta}$ are 
expressible in terms of Clausen functions (see Appendix~A of Ref.~\cite{DK01}.)
To obtain a result valid in a different region of variable $u_\theta \; (-4 \leq u_\theta \leq 0)$,
the analytical continuation procedure, firstly described in Ref.~\cite{bastei} (see also \cite{DK01}) 
should be applied.

{\bf 5.}
Our study \cite{DK04} allows us to construct some terms of the 
$\ep$-expansion of the generalized hypergeometric function ${}_{P+1}F_P$
and obtain new analytical results for higher terms of the $\ep$-expansion
of the  two- and three-loop maser-integrals entering in different packages 
\cite{onshell2,leo96}. 

It was noted, that individual terms of our construction 
include the log-sine functions of arguments $\theta/2$ and $\pi-\theta/2$ 
(see Eqs.~(\ref{rho3:res}, \ref{SIGMA3}) and Eq.~(2.90) in Ref.~\cite{DK04})
which, however, are cancelled  in the considered order of $\ep$-expansion of the 
Feynman diagrams.  For single scale massive diagrams, when $\theta = \pi/3$, 
the arguments of functions are equal to $\pi/6$ and $5\pi/6$, respectively.
It opens the question about possible generalization of the 
``sixth root of unity'' basis, introduced in Ref.~\cite{sixth_root_of_unity}
(for the recent progress see \cite{DK01,even:odd}).

\noindent
{\bf Acknowledgements.}
I would like to thank the organizers of `Loops and Legs 2004' conference.
My special thanks are to A.~Davydychev and F.~Jegerlehner for collaboration;
discussion with O.~Tarasov was quite useful. I indebted also to 
R.~Bonciani, D.~Broadhurst, M.~Hoffmann, Hoang Ngoc Minh, 
G.~Passarino, E.~Remiddi and V.~Smirnov for interesting discussion.
I thank G.~Sandukovskaya for careful reading of the manuscript. 

\end{document}